\documentclass[a4paper,11pt]{article}
\pdfoutput=1 

\usepackage{jinstpub} 

\usepackage{amsmath}

\usepackage{float}

\title{\boldmath Longevity Study on the CMS Resistive Plate Chambers for HL-LHC}

\author[n,1]{R. Aly \note{Corresponding author.}}
\author[]{\newline}
\author[]{\\ on behalf of the CMS Collaboration }

\affiliation[n]{INFN, Sezione di Bari, Via Orabona 4, IT-70126 Bari, Italy.}

\emailAdd{reham.aly@cern.ch}

\abstract{The CMS Resistive Plate Chamber (RPC) system has been certified for 10 years of LHC operation. In the next years, during the High luminosity LHC (HL-LHC) phase, the LHC instantaneous luminosity will increase to a factor five more than the existing LHC luminosity. This will subject the present CMS RPC system to background rates and operating conditions much higher with respect to those for which the detectors have been designed. Those conditions could affect the detector properties and introduce  nonrecoverable aging effects. A dedicated longevity test is set up in the CERN Gamma Irradiation Facility (GIF++) to determine if the present RPC detectors can survive the hard background conditions during the HL-LHC running period. During the irradiation test, the RPC detectors are exposed to a high gamma radiation for a long period and the detector main parameters are monitored as a function of the integrated charge. Based on collecting a large fraction of the expected integrated charge at the LH-LHC, The results of the irradiation test will be presented.}

\keywords{Gas detectors, Aging, Resistive Plate Chamber, CMS, HL-LHC }

\proceeding{The 12$^{\text{th}}$ International Conference on Position Sensitive Detectors\\
12 -17 September, 2021\\
 Birmingham, United Kingdom}

\begin{document}
\maketitle
\flushbottom

\section{Resistive plate chambers at CMS}
\label{sec:RPC}

The Resitive Plate Chambers (RPCs) at the Compact Muon Solenoid (CMS) experiment at the CERN Large Hadron Collider (LHC) provide redundancy to the muon trigger system and contributes to the muon reconstruction and identification \cite{cms-detector,cms}. In CMS the RPC system consists of 1056 chambers installed in the barrel and endcap regions covering a pseudo rapidity region $|\eta|$ = 1.9 \cite{cms,muon-system}. The RPC chamber consists of two layers of 2 mm gas gaps with a sheet of segmented copper readout strips placed between them. Each gas gap is made of two sheets of High-Pressure-Laminate (HPL) electrodes with 2 mm thickness and filled with a nonflammable three-component gas mixture of 95.2\% freon (C${_2}$H${_2}$F${_4}$), 4.5\% isobutane (i-C${_4}$H$_{10}$) and 0.3\% sulphur hexafluoride (SF${_6}$).

\section{Motivation of Aging study}
\label{sec:motivation}

When a gas detector is exposed to high radiation for long time, it can suffer from aging effects which result in a degradation of detector performance appearing as loss in detector efficiency, increase in dark current\footnote{Dark current is the current produced in the chamber when applying high voltage in the absence of background radiation.} and rise in noise rates. 
The RPC system worked efficiently during the LHC Runs I and II data taking periods at the nominal luminosity of 1 $\times$ 10$^{34}$ cm$^{-2}$ s$^{-1}$ without any degradation of the detector performance \cite{muon-performanceRunI,muon-performance1,muon-performance2}. During HL-LHC phase, the LHC instantaneous luminosity will increase to a factor five more than the nominal LHC luminosity. Based on the data collected by CMS during the LHC Run II, and assuming a linear dependence of the background rates as a function of the instantaneous luminosity, the expected background rates and integrated charge at HL-LHC (including a safety factor of three) will be about 600  Hz/cm$^2$ and 840 mC/cm$^2$, respectively \cite{HL-LHC}. Nonrecovable aging effect can appear in those operating condition, that can affect the detector performance and properties. To determine whether the present RPC detectors can survive the hard background conditions during the HL-LHC running period, a long term irradiation test has been carried out at the CERN Gamma Irradiation Facility (GIF++) since July 2016\cite{gif++}.


\section{Set-up and procedure}
\label{sec:setup}

\begin{figure}[htbp]
\centering 
\includegraphics[width=.3\textwidth,origin=c]{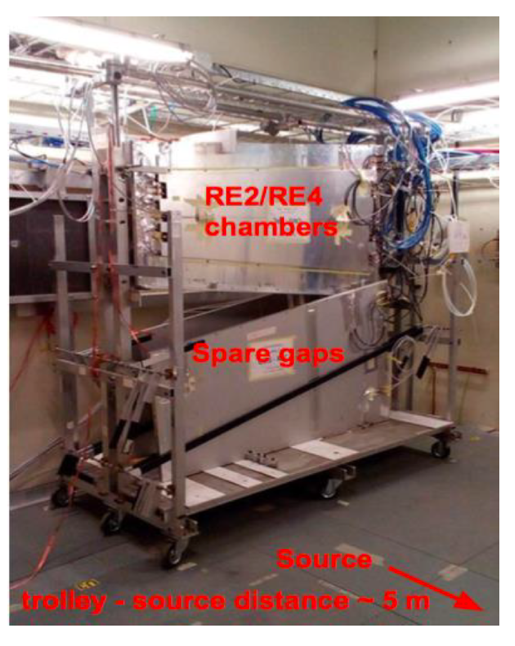}
\includegraphics[width=.5\textwidth,origin=c]{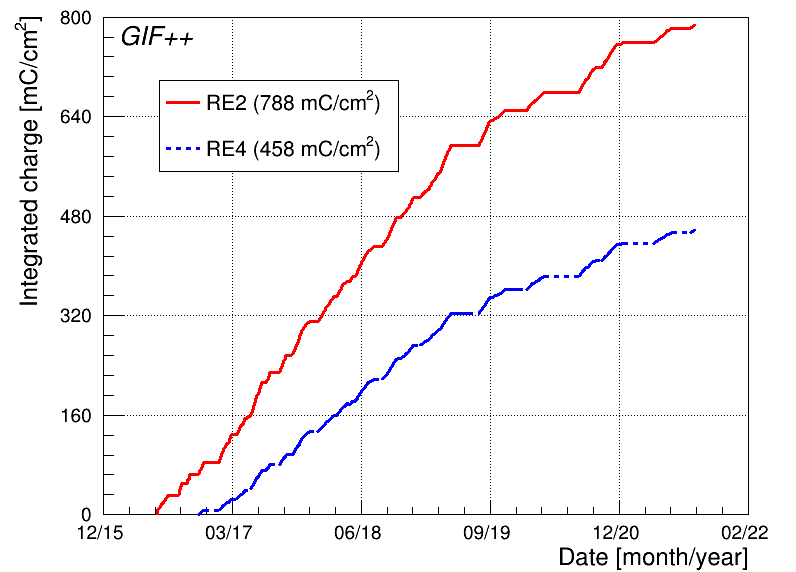}
\caption{\label{fig:qint} Longevity test setup in GIF++ (left) and Integrated charge versus time, accumulated during the longevity test at GIF++ for RE2/2 (solid red line) and RE4/2 (dashed blue line) chambers (right). The RE4/2 chamber has been turned on a few months later because of total gas flow limitations. Different slopes account for different irradiation conditions during data taking.}
\end{figure}

The irradiation test is ongoing on four endcap spare RPC chambers, two RE2/2 and two RE4/2 chambers, as shown in Fig.\ref{fig:qint}(left). Two chambers (one RE2 and one RE4) are continuously under irradiation, while the other two chambers of the same type are kept as a reference. The reference chambers are switched on from time to time to compare the results with the irradiated chamber. 
The detector parameters (such as dark current, noise rate, current and count rates at several background conditions) are monitored continuously and compared with the measurements from the reference chambers to identify any degradation in the irradiated chambers due to long term irradiation. Moreover, when the muon beam at GIF++ is available, the detector performance is measured at various irradiation fluxes. The integrated charge\footnote {The integrated charge is calculated as the average density current accumulated in time in the three gaps that constitute the detector.} collected from the beginning of irradiation is about 788 and 458 mC/cm$^2$ for RE2 and RE4 chambers respectively, as shown in Fig.\ref{fig:qint}, that correspond to approximately 93 $\%$  and 54 $\%$ of the expected integrated charges at HL-LHC.

\section{Detector parameter monitoring}

\begin{figure}[htbp]
\centering 
\includegraphics[width=.32\textwidth,origin=c]{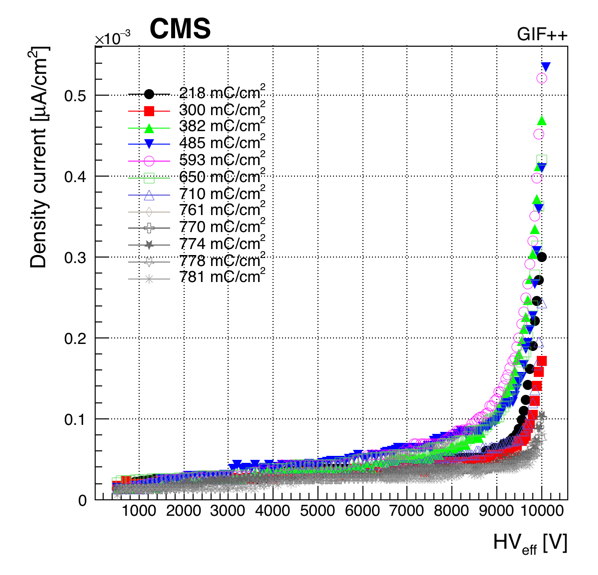}
\includegraphics[width=.32\textwidth,origin=c]{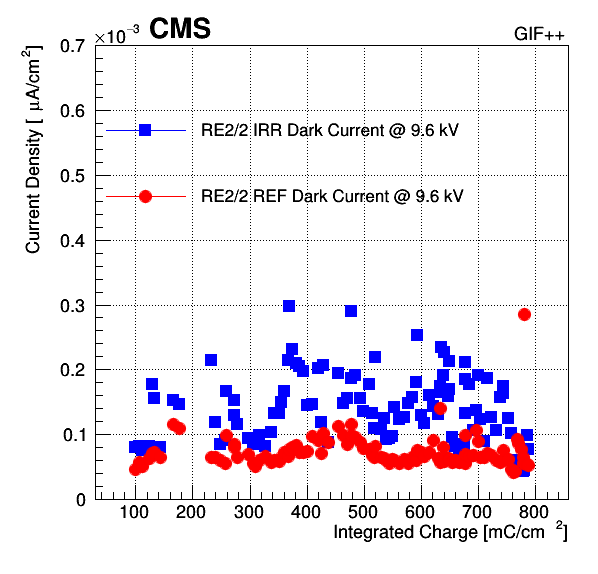}
\includegraphics[width=.32\textwidth,origin=c]{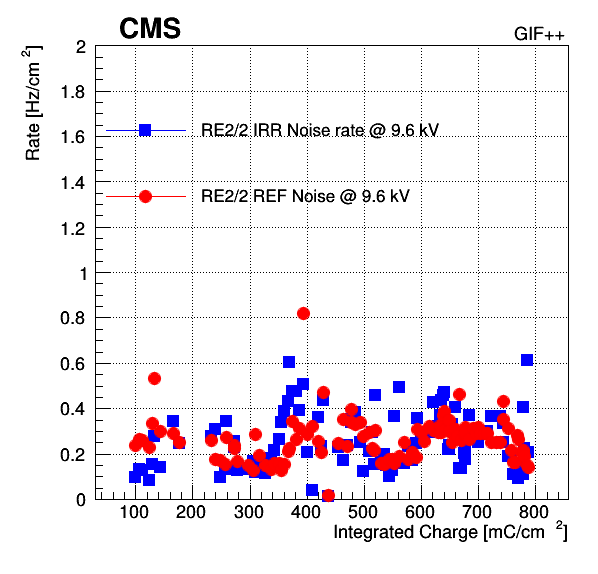}
\caption{\label{fig:hv_scan} Dark current density, monitored as a function of the effective high voltage at various values of collected integrated charge for the RE2 irradiated chamber (left). Dark current density for the RE2 irradiated (blue squares) and the reference (red circles) chambers as a function of collected integrated charge at 9.6 kV (middle) and the average noise rate as a function of collected integrated charge for the RE2 irradiated (blue squares) and the reference chambers (red circles) (right).}
\end{figure}

During the irradiation process, the dark current and noise rate are monitored periodically to identify any aging effects due to irradiation. Fig. \ref{fig:hv_scan} (left) shows the dark current density for RE2 irradiated chamber, monitored as a function of effective high voltage\footnote{Effective high voltage is the voltage normalized at the standard temperature  20 $^o$C and pressure 990 $mbar$ \cite{HV_Norm}} at various values of collected integrated charge. Also, the dark current density at 9.6 kV for RE2 both irradiated (blue) and reference (red) chambers as a function of collected integrated charge is shown in Fig. \ref{fig:hv_scan} (middle), where the dark currents were measured at 9.6 kV, which includes the gas amplification contribution. The dark current is almost stable in time with small acceptable variations since the beginning of irradiation.
The average noise rate for the RE2 irradiated (blue) and reference (red) chamber is measured as a function of collected integrated charge, as shown in Fig. \ref{fig:hv_scan} (right). The average noise rate is stable with time and less than 1 Hz/cm$^2$.

\section{Detector performance }

\begin{figure}[htbp]
\centering 
\includegraphics[width=.45\textwidth,origin=c]{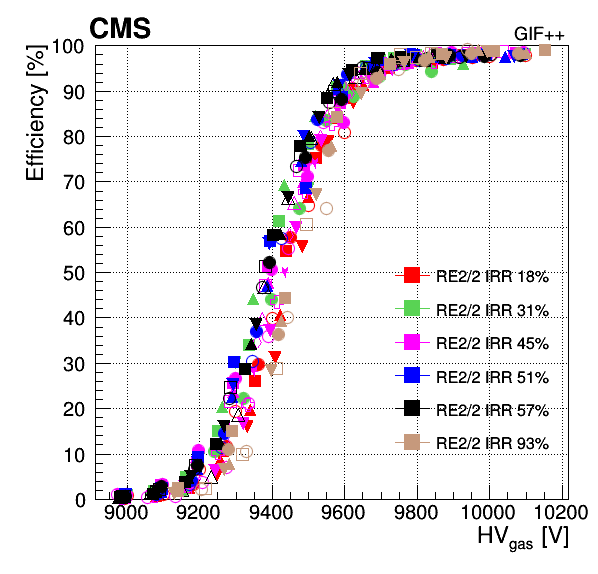}
\includegraphics[width=.45\textwidth,origin=c]{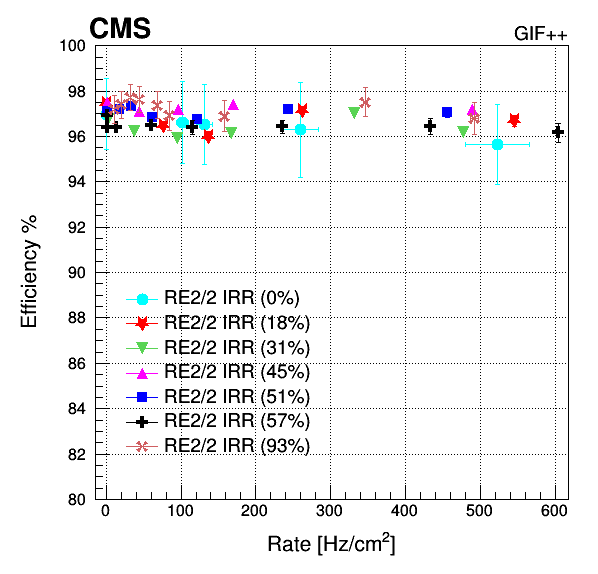}
\caption{\label{fig:hvgas} RE2/2 irradiated chamber efficiency as a function of the HV$_{gas}$ (left) at various background irradiation rates (various marker shapes with same color) and various integrated charge values. The RE2 irradiated chamber efficiency at working point as a function of the background rate at various values of collected integrated charge (right) .}
\end{figure}

When the muon beam is present in GIF++, the detector performance has been measured at various periods of irradiation. Fig. \ref{fig:hvgas} (left) shows the efficiency measured at various values of collected integrated charge and at various background radiation rates up to 600 Hz/cm$^2$. At high background rates, the effective voltage applied to the electrodes and the effective voltage on the gas HV$_{gas}$ are not the same due to the voltage drop across the electrodes \cite{mypaper,HV_Norm2}. So, the detector efficiency is measured as a function of the HV$_{gas}$ to exclude any effect related to the voltage drop across the electrodes since the detector operation regime is invariant with respect to HV$_{gas}$. The efficiency is stable in time and  compatible with the values in the beginning of irradiation.
The RE2 irradiated chamber efficiency at working point is measured at various background rates (up to 600 Hz/cm$^2$) and at various integrated charge values, as shown in Fig.\ref{fig:hvgas} (right).
The efficiency is stable in time up to the highest background rate expected at HL-LHC (600 Hz/cm$^2$ ). This result demonstrates the stability of the detectors efficiency, and that they can operate efficiently at the maximum expected rate during HL-LHC.

\begin{figure}[htbp]
\centering 
\includegraphics[width=.45\textwidth,origin=c]{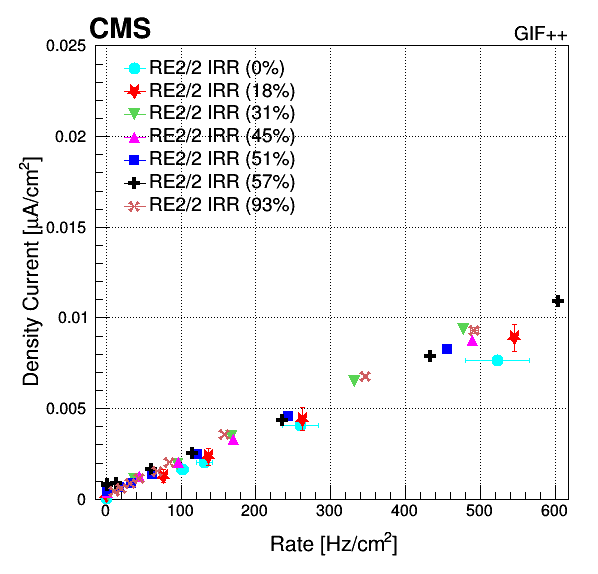}
\includegraphics[width=.45\textwidth,origin=c]{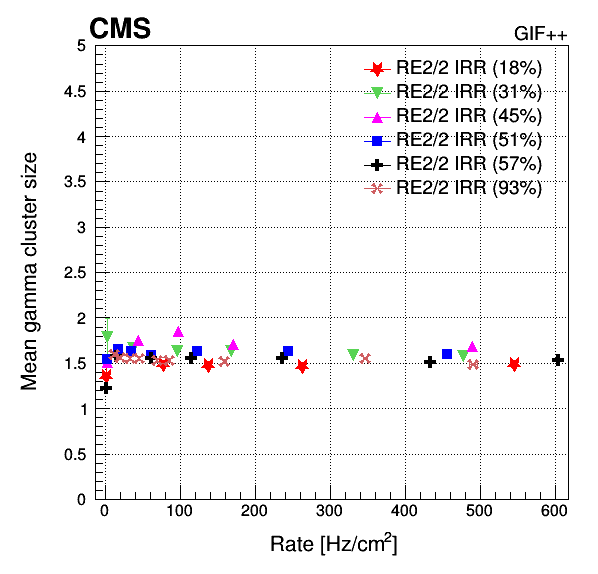}
\caption{\label{fig:plots_at_wp} RE2/2 irradiated chamber density current (left) and gamma cluster size (right) measured at the detector working point as a function of the background rate, at various integrated charge values.}
\end{figure}

Fig. \ref{fig:plots_at_wp} shows the current density (left) and gamma cluster size
\footnote{The number of contiguous strips which give signals at the crossing of an ionizing particle.} (right) at the detector working point, measured for the RE2 irradiated chambers at various values of integrated charge up to 600 Hz/cm$^2$. The current and the cluster size are stable from the beginning of irradiation and the cluster size is $<$2. 

\section{Conclusion}
Longevity studies on spare RPCs are ongoing at GIF++ under controlled conditions. 
 Results show that the RPC chambers performance and parameters are stable up to 93$\%$ of the expected integrated charge up to 600 Hz/cm$^2$. We conclude that the current RPC system is capable of reliable operation in the HL-LHC up to the given amounts of integrated charge.

\acknowledgments

We would like to thank our colleagues from CERN Gamma Irradiation facility where the measurements leading to those results have been performed. Also my sincere thanks to all CMS RPC members for their valuable work and PSD12 organizers for a great conference.

\end{document}